\documentclass[12pt]{article}
\usepackage{graphicx}

\newcommand\pubnumber{}
\newcommand\pubdate{\today}

\def\harvard{Department of Physics\\
Harvard University}

\def\Title#1{\begin{center} {\Large #1 } \end{center}}
\def\Author#1{\begin{center}{ \sc #1} \end{center}}
\def\Address#1{\begin{center}{ \it #1} \end{center}}

\newcommand\pubblock{\rightline{\begin{tabular}{l} \pubnumber\\
         \pubdate  \end{tabular}}}
\newenvironment{Abstract}{\begin{quotation}  }{\end{quotation}}
\newenvironment{Presented}{\begin{quotation} \begin{center} 
             PRESENTED AT\end{center}\bigskip 
      \begin{center}\begin{large}}{\end{large}\end{center} \end{quotation}}

\def\beq{\begin{equation}}
\def\eeq#1{\label{#1}\end{equation}}
\def\eeqn{\end{equation}}

\def\beqa{\begin{eqnarray}}
\def\eeqa#1{\label{#1}\end{eqnarray}}
\def\eeqan{\end{eqnarray}}

\let\bar=\overbar

\def\Dslash{\not{\hbox{\kern-4pt $D$}}}
\def\dslash{\not{\hbox{\kern-2pt $\del$}}}

\def\BR{\mbox{\rm BR}}

\def\msb{{\bar{\ssstyle M \kern -1pt S}}}

\input babarsym
\newcommand{\bfig}{\begin{figure}[htbpc!]}
\newcommand{\efig}{\end{figure}}

\providecommand{\etapr}{\ensuremath{\eta^\prime}}

\newcommand{\fetaprhoz}{\ensuremath{\etapr\rho^0}}
\newcommand{\fetaprhop}{\ensuremath{\etapr\rho^+}}

\providecommand{\KS}{\ensuremath{K_S^0}}

\newcommand{\rhoz}{\ensuremath{\rho^0}}
\newcommand{\rhop}{\ensuremath{\rho^+}}

\def\qqbar{\ensuremath{q\bar q}}

\def\beq{\begin{equation}}
\def\eeq{\end{equation}}
\def\bef{\begin{figure}}
\def\edf{\end{figure}}
\def\ben{\begin{enumerate}}
\def\een{\end{enumerate}}
\def\bear{\begin{array}}
\def\enar{\end{array}}
\def\beqa{\begin{eqnarray}}
\def\eeqa{\end{eqnarray}}
\def\bit{\begin{itemize}}
\def\eit{\end{itemize}}

\def\to{\ensuremath{\rightarrow}}

\def\mes{\ensuremath{{m_{ES}}}}

\newcommand{\Kst}{\ensuremath{K^{*}}\xspace}
\newcommand{\Kstz}{\ensuremath{K^{*0}}\xspace}

\newcommand{\Kstp}{\ensuremath{K^{*+}}\xspace}

\newcommand{\KstTwo}{\ensuremath{K_2^{*}(1430)}\xspace}

\renewcommand{\pipi}{\ensuremath{\pi\pi}\xspace}

\newcommand{\ems}{\ensuremath{\times 10^{-6}}\xspace}
\newcommand{\fetapfz}{\ensuremath{\etapr f_0}\xspace}
\newcommand{\fetapKstz}{\ensuremath{\etapr\Kstz}}
\newcommand{\fetapKstp}{\ensuremath{\etapr\Kstp}}
\newcommand{\fzero}{\ensuremath{f_0}}
\newcommand{\KpiSwave}{\ensuremath{(K\pi)^*_0}}

\begin{document}
\begin{titlepage}
\pubblock

\vfill
\Title{Experimental Results in Charmless Hadronic $B$ Decays from the $B$ Factories}
\vfill
\Author{Corry L. Lee}
\Address{\harvard}
\vfill
\begin{Abstract}
We report on recent measurements, from the \babar\ and Belle experiments, of $B$-meson decays to purely hadronic final states that do not contain charm.  The studies are based on large samples of \BB\ pairs collected at the \FourS\ or $B_s^{(*)}\overline{B}_s^{(*)}$ pairs collected at the \FiveS\ by the \babar\ and Belle detectors at the asymmetric energy \epem\ colliders at SLAC and KEK-B, respectively.  This paper includes the following results: measurements of branching fractions and charge asymmetries of $B$ meson decays to $\eta^{\prime}\rho$, $\eta^{\prime} f_{0}$, and $\eta^{\prime} K^*$, where the $K^*$ stands for a vector, tensor, or scalar strange meson; a search for $\Bz\ra\Kp\pim\Kmp\pipm$, including the \Kstarz\ resonance; a search for $\Bp\ra a_1^+ K^{*0}$, an axial-vector vector final state; a measurement of $B_s^0\ra hh$ branching fractions, where $h=\Kp,\KS,\mbox{or }\pip$; and inclusive branching fraction measurements of $\Bp\ra\Kp\piz\piz$ and $\Bz\ra\pip\KS\Km$.  
\end{Abstract}
\vfill
\begin{Presented}
The 6th International Workshop on the CKM Unitarity Triangle, \\
University of Warwick, UK, 6-10 September 2010
\end{Presented}
\vfill
\end{titlepage}
\def\thefootnote{\fnsymbol{footnote}}
\setcounter{footnote}{0}

\section{Introduction}

Experimental studies of charmless hadronic $B$ decays provide a strong test of theoretical calculations and serve as a laboratory in which to search for potential new physics effects.  New physics effects can arise from new particles and couplings in the loop diagrams through which many of these decays proceed.  Identifying new physics effects requires a solid theoretical description of Standard Model (SM) processes, which is complicated by the interplay of long- and short-distance QCD effects.  Many theoretical predictions have been made by Perturbative QCD (pQCD), QCD Factorization (QCDF), Soft Colinear Effective Theory (SCET), and Na\"ive Factorization (NF), though often with large uncertainties~\cite{JimCheng}.

We report on recent results from the \babar\ and Belle Collaborations.  Where appropriate, we compare the results to theoretical predictions and previous measurements.  Charge conjugate states are implied throughout this paper, and all upper limits are quoted at the 90\% confidence level.

The \babar\ results described below use the full data sample of roughly $470\times 10^6$ \BB\ pairs; the Belle results use $657\times 10^6$ \BB\ pairs.  These datasets allow access to branching fractions (BF) $\sim 10^{-5}-10^{-7}$.  The Belle measurement of $B_s^0\ra hh$ is made on $1.25\times 10^6$ $B_s^{(*)}\overline{B}_S^{(*)}$ pairs collected at the \FiveS.  

Experimental studies of charmless hadronic $B$ decays employ maximum likelihood fits to discriminate between signal, the dominant background from $\epem\ra\qqbar$ (where $q=u,d,s,c$), and backgrounds from other $B$ decays.  The analyses typically make use of the fully reconstructed final state to define two primary observables:  \mes\ (\babar) or $M_{bc}$ (Belle), which uses the beam energy and reconstructed final state momentum to define an observable which peaks at the $B$ mass~\cite{PDG} for signal; and \DeltaE, the energy difference between the reconstructed $B$ candidate and the beam energy.   Additional background discrimination is provided by event shape variables, as the decay products of $B$ decays tend to be spherically distributed in the \FourS\ rest frame, while the \qqbar\ background is jet-like.  Event shape variables are combined into a Fisher discriminant or artificial neural network (NN).  Where appropriate, resonance mass and helicity distributions are also included in the maximum likelihood fit.

\section{Experimental Results}

Recent experimental results from \babar\ and Belle are presented in this section.

\subsection{$B$ meson decays to $\eta^{\prime}\rho$, $\eta^{\prime} f_{0}$, and $\eta^{\prime} K^*$}

\babar\ measures branching fractions and, where appropriate, charge asymmetries for $B$ meson decays to $\eta^{\prime}\rho$, $\eta^{\prime} f_{0}$, and $\eta^{\prime} K^*$, where the $K^*$ stands for a vector $K^*(892)$, tensor $K_2^*(1430)$, or scalar $K_0^*(1430)$ interfering with the non-resonant scalar $K\pi$~\cite{etapV}.  Such measurements test predicted $\eta/\etapr$ mixing as well as provide access to potential new physics effects.  Theoretical predictions for the $\Bp\ra\etapr\rhop$ branching fraction in pQCD and QCDF ($6-9\times 10^{-6}$) disagree with those from SCET ($\sim 0.4\times 10^{-6}$).  Few theoretical predictions exist for decays involving the $K_2^*(1430)$ or $K_0^*(1430)$.

\begin{figure}[htb]
\centering
  \includegraphics[width=1.0\linewidth]{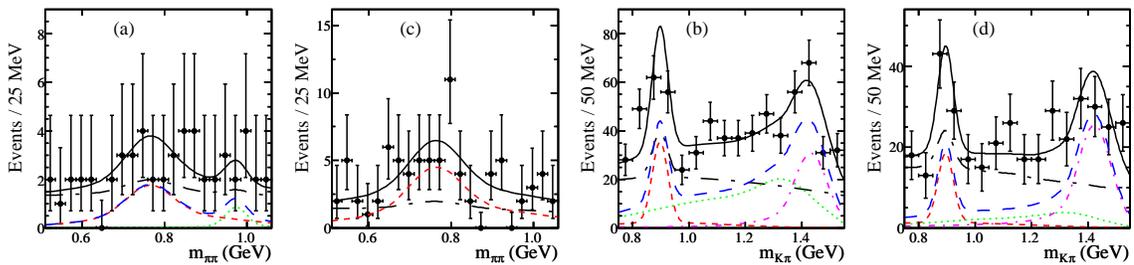}
    \caption{$B$-candidate $\pi\pi$ invariant mass projections for: (a) 
    \fetaprhoz/\fetapfz, (b) \fetaprhop; and $K\pi$ invariant mass for: (c) 
    \fetapKstz, (d) \fetapKstp. The solid curve is the fit 
    function, black long-dash-dot is the total background, 
    and the blue dashed curve is the total signal contribution.  
    In (a) the \rhoz\ component (red dashed) is separated from 
    the \fzero\ (green dotted). In (c,d), the $K^*(892)$ (red dashed) is separated 
    from the \KpiSwave\ (green dotted) and the \KstTwo\ 
    (magenta dot-dashed) components.}
  \label{fig:etapKst}
\end{figure}

First observations are presented for four modes, including $\Bp\ra\etapr\rhop$, which has a measured branching fraction of $(9.7^{+1.9}_{-1.8}\pm 1.1)\times 10^{-6}$.  This measurement favors the pQCD and QCDF predictions and is in poor agreement with the upper limit from Belle ($<5.8\times 10^{-6}$)~\cite{BelleEtapRho}.  The branching fractions involving the tensor \KstTwo are substantially higher than those involving the $K^*(892)$, a pattern also observed in $B\ra\omega\Kst$ decays~\cite{omegaKst}.
Plots of the $\pi\pi$ and $K\pi$ invariant masses are shown in Fig.~\ref{fig:etapKst}, including a cut on the likelihood function to enhance the visibility of signal.

\subsection{Search for $\Bz\ra\Kp\pim\Kmp\pipm$}

\begin{figure}[htb]
\centering
  \scalebox{1.0}[1.0]{\includegraphics[width=0.33\linewidth]{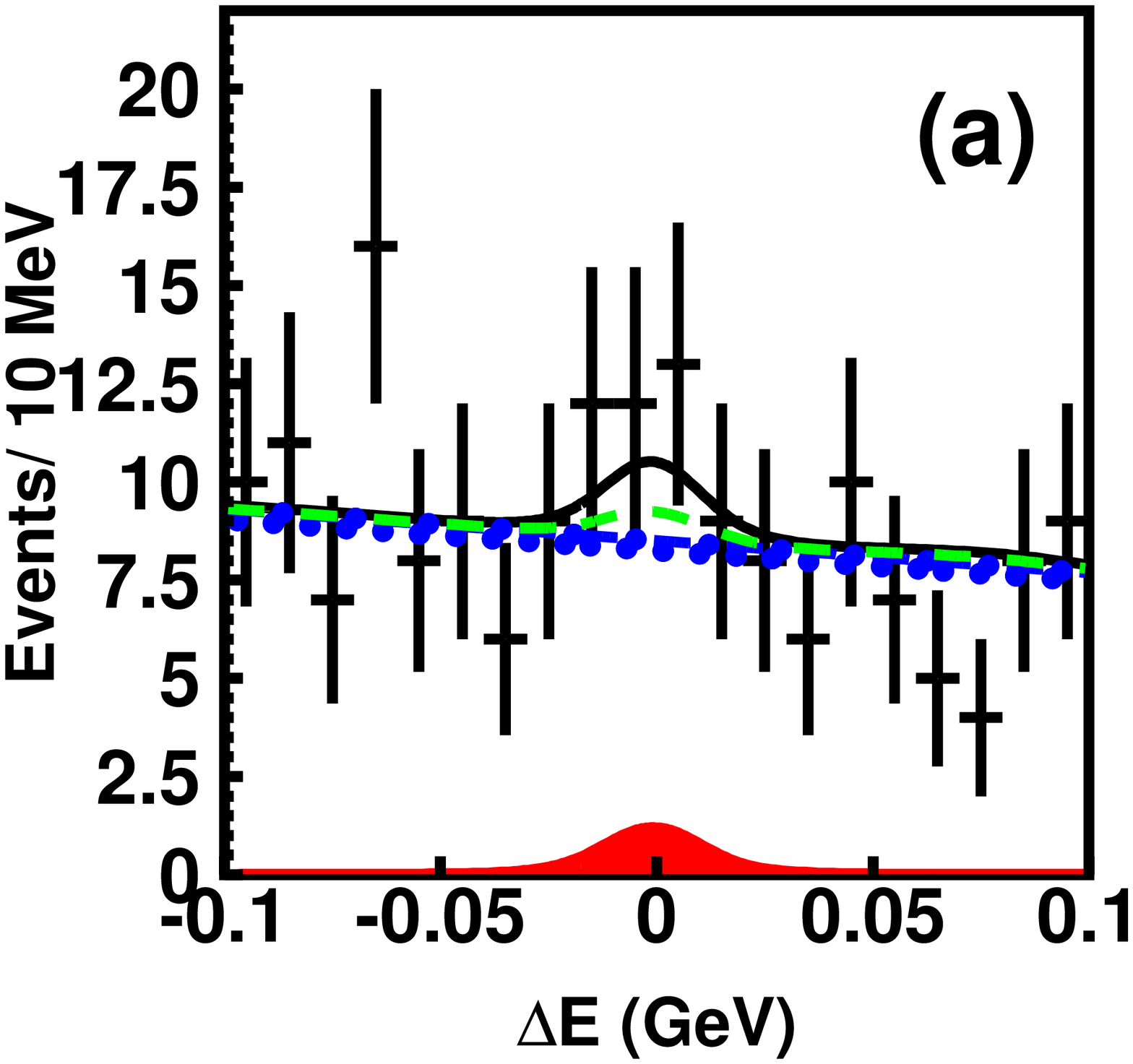} \includegraphics[width=0.33\linewidth]{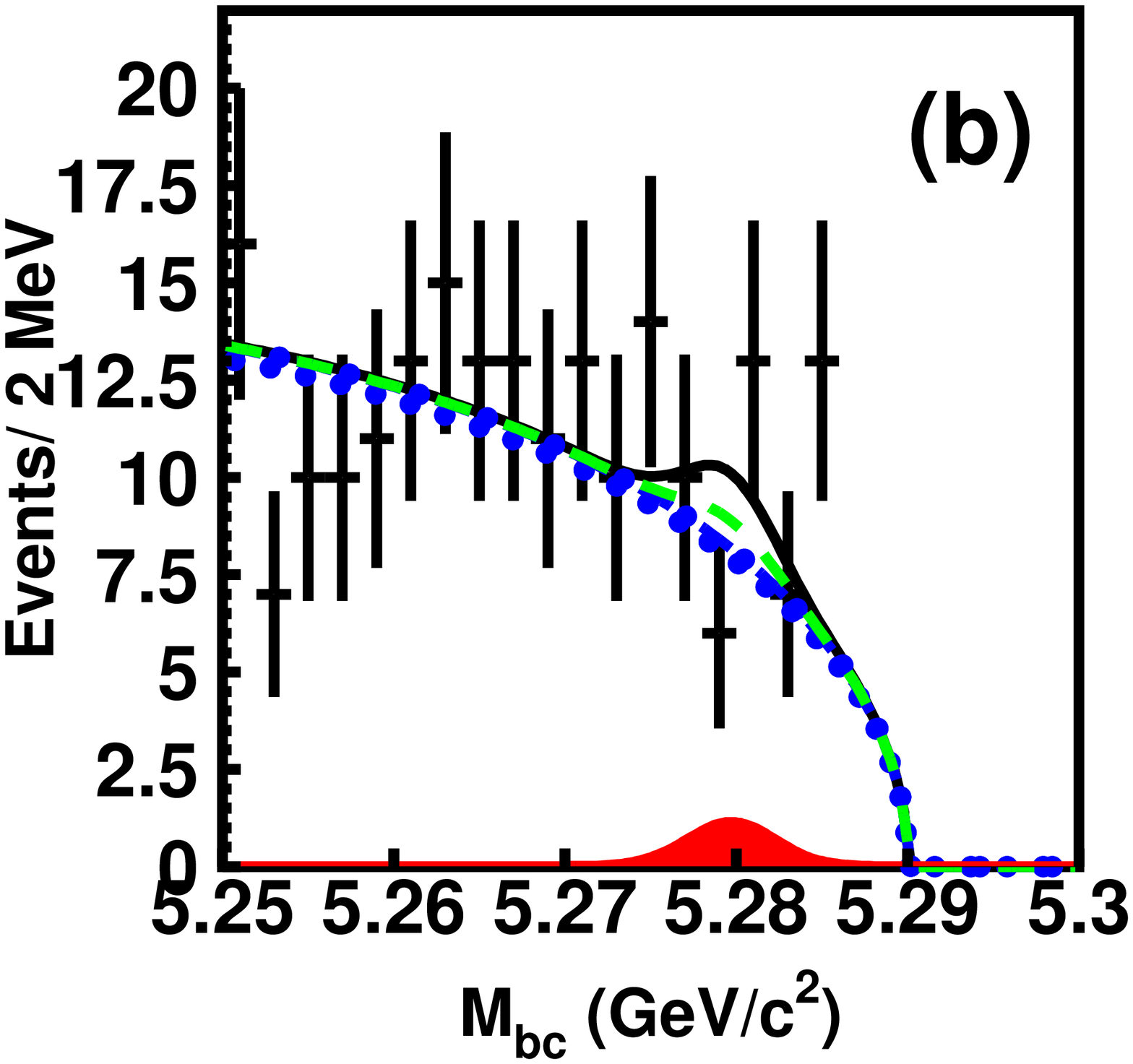} \includegraphics[width=0.33\linewidth]{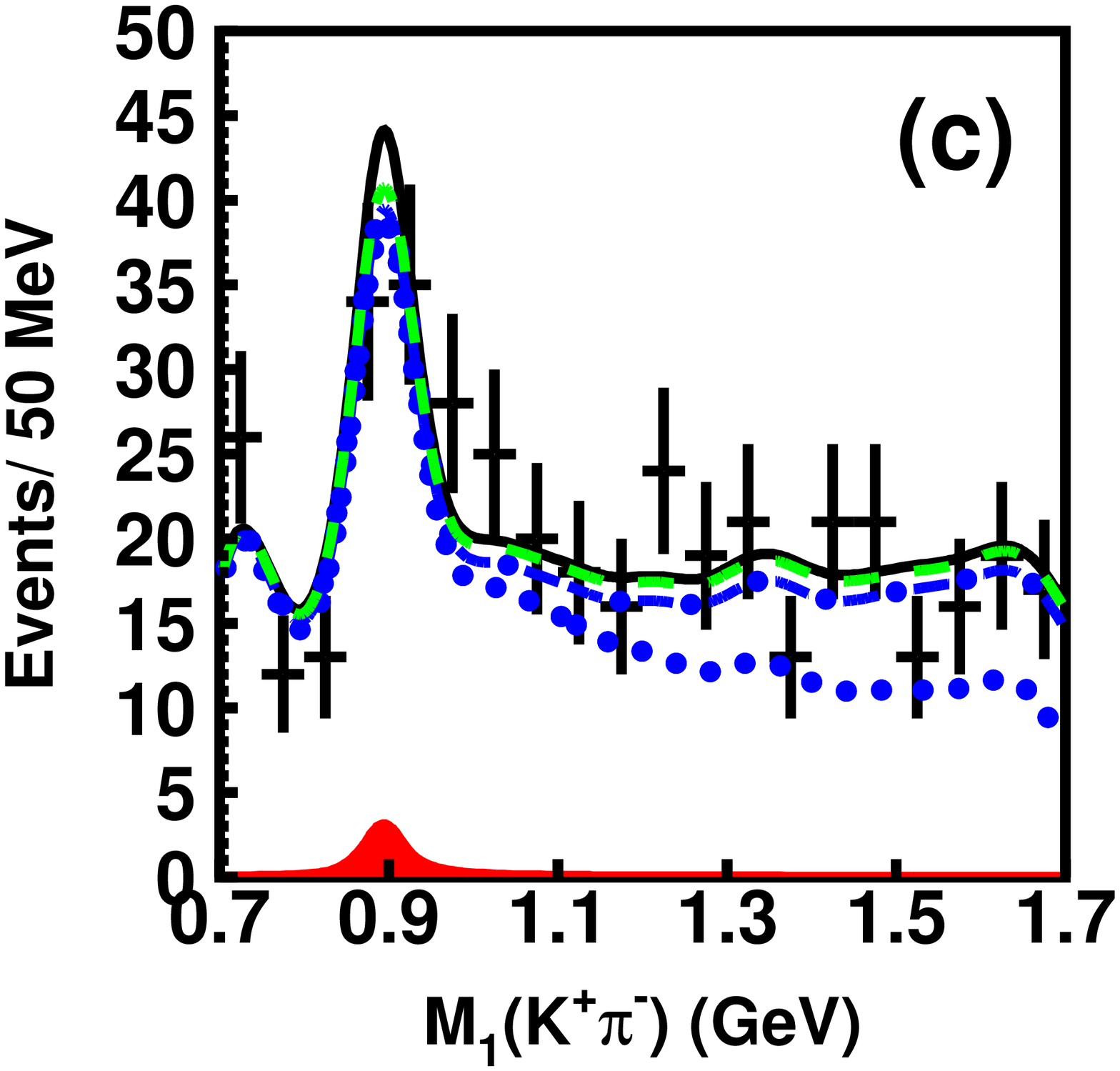}}
    \caption{Projections of (a) \DeltaE, (b) $M_{bc}$, and (c) $\Kp\pim$ invariant mass for $\Bz\ra\Kstarz\Kstarzb$.  The black curve shows the overall fit result; the red shaded region represents the signal component; the blue dotted curve represents \qqbar\ background, the dot-dashed blue is $b\ra c$ background and the green dashed is $B\ra$ charmless background.
    }
  \label{fig:KstKstb}
\end{figure}

Belle searches for $\Bz\ra\Kp\pim\Kmp\pipm$ where a $K\pi$ pair can be a resonant \Kstz\ or \Kstarzb, where \Kstz\ represents either the vector $K^*(892)^0$ or scalar $K_0^*(1430)^0$~\cite{belleKstKst}.  The $\Bz\ra\Kstz\Kstarzb$ decay is dominated by a $b\ra d$ penguin, and is expected to have a BF of $\sim 10^{-7}-10^{-6}$ in the SM.  The decay $\Bz\ra\Kstz\Kstz$ is suppressed in the SM, the expected BF $\sim 10^{-15}$.  No significant signals are observed.  BF upper limits are placed:  BF$(\Bz\ra\Kstz\Kstarzb)< 0.8\ems$ and BF$(\Bz\ra\Kstz\Kstz)< 0.2\ems$.  The former limit is slightly below the \babar\ measurement of BF$(\Bz\ra\Kstz\Kstarzb)=(1.28\pm 0.34)\ems$~\cite{babarKstKst}.  Projections of the $\Kstz\Kstarzb$ fit results on the signal-enhanced data sample are given in Fig.~\ref{fig:KstKstb}.

\subsection{Search for $\Bp\ra a_1^+(1260) K^{*0}$}

\babar\ presents a search for the axial-vector vector decay $\Bp\ra a_1^+(1260) K^{*0}(892)$~\cite{a1Kst}.  Theoretical predictions of the BF differ greatly between methods, QCDF predicting $\sim 11\ems$ and NF calculating a BF an order of magnitude smaller.  No significant signal is observed, and \babar\ sets a 90\% confidence level upper limit on the BF $<3.6\ems$, assuming an equal BF for $a_1^+(1260)\ra\pip\pim\pip$ and $a_1^+(1260)\ra\pip\piz\piz$, and that the BF for $a_1^+(1260)\ra 3\pi$ is 100\%.

\subsection{$B_s^0\ra hh$, where $h=\Kp,\KS,\mbox{or }\pip$}

Belle presents results for $B_s^0\ra hh$, where $h=\Kp,\KS,\mbox{or }\pip$~\cite{BelleBs}.  Understanding these channels could help understand the ``$K\pi$ puzzle''~\cite{JimCheng} in $B^0$ decays, and comparing charge asymmetries between $B$ and $B_s$ decays could provide a window on new physics.  Belle measures BF$(B_s^0\ra\Kp\Km)=(38\pm12)\ems$, in good agreement with (though with larger errors than) CDF~\cite{CDFBs}, and presents upper limits on $B_s^0\ra\Kp\pim,\ \pip\pim,\mbox{ and }\Kz\Kzb$.  This is the first search for $B_s^0\ra\Kz\Kzb$, and Belle places the upper limit of $< 66\ems$.

\subsection{Inclusive $\Bp\ra\Kp\piz\piz$}

\begin{figure}[htb]
\centering
    \scalebox{1.0}[1.0]{\includegraphics[width=0.5\linewidth]{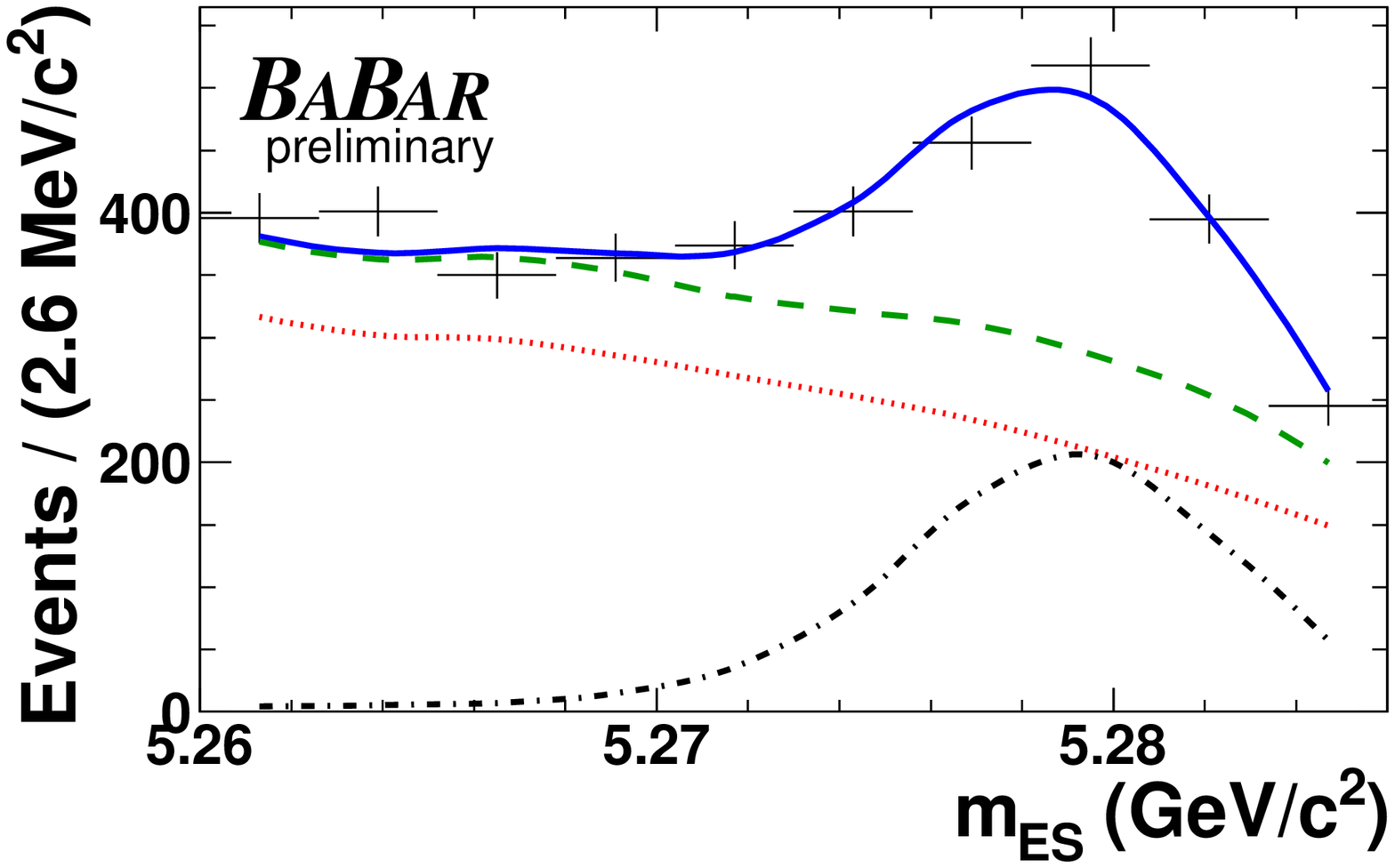} \includegraphics[width=0.5\linewidth]{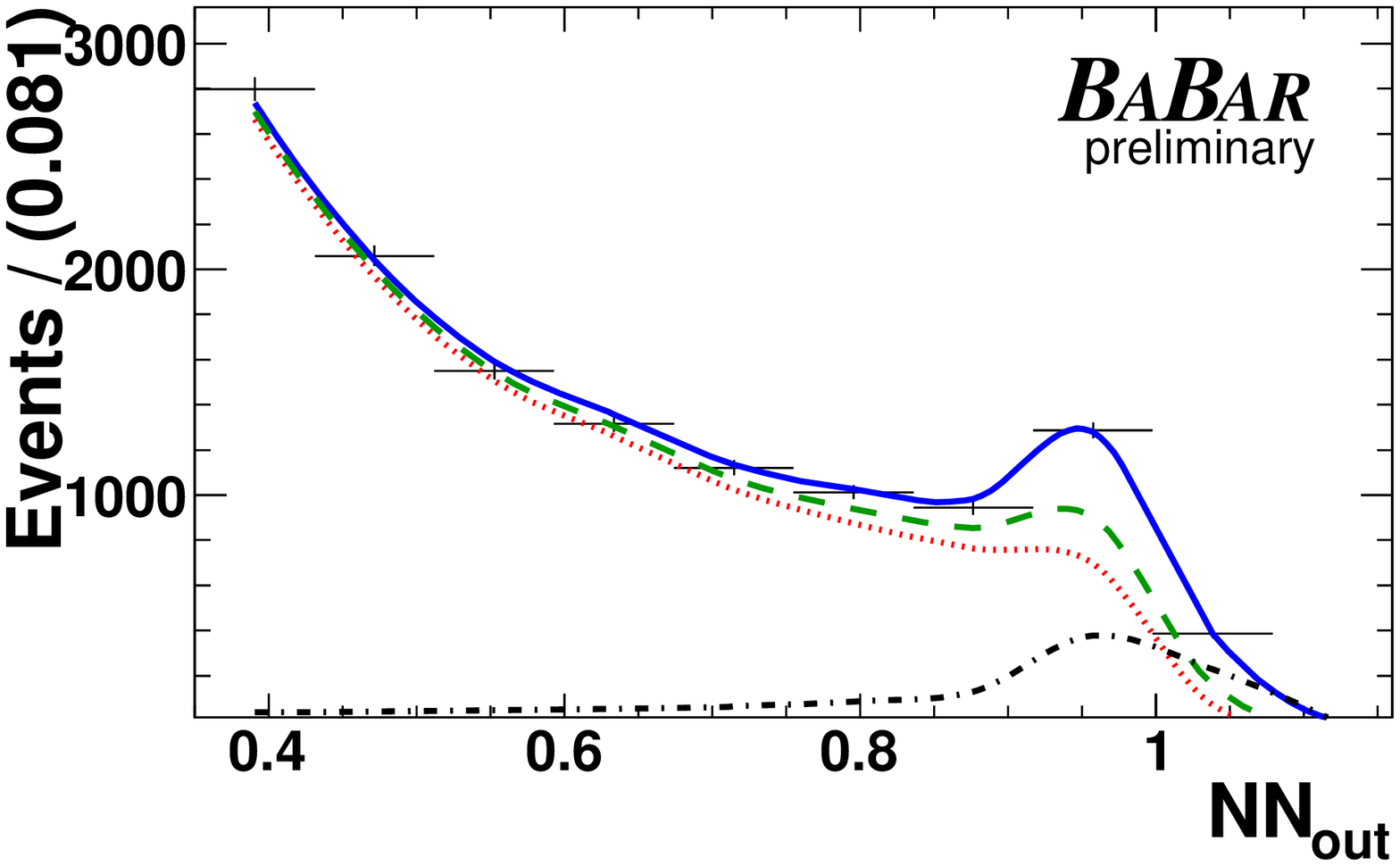}}
    \caption{Projections of $\Bp\ra\Kp\piz\piz$ candidates onto (a) \mes\ and (b) neural network, including cuts to enhance signal visibility.  The solid blue curve represents the total fit result, the dot-dashed black curve gives the signal contribution, the red dotted curve the \qqbar, and the green dashed curve the total background contribution.
    }
  \label{fig:Kpi0pi0}
\end{figure}

\babar\ reports a preliminary measurement of the inclusive $\Bp\ra\Kp\piz\piz$ branching fraction~\cite{Kpi0pi0}.  Understanding $B\ra\Kst\pi$ decays could help shed light on the ``$K\pi$ puzzle''~\cite{JimCheng}, and $\Bp\ra\Kstp\piz$ is poorly measured, with the three-body state having never been previously investigated.  This analysis employs \mes\ and a event-shape neural network discriminant, yielding $1220\pm 85$ signal events.  The measured BF is $(15.5\pm 1.1\pm 1.6)\ems$ with a significance $>10\sigma$.  Projections of the fit results on the signal-enhanced data sample are given in Fig.~\ref{fig:Kpi0pi0}.

\subsection{Inclusive $\Bz\ra\pip\KS\Km$}

\babar\ observes the inclusive $\Bz\ra\pip\KS\Km$ decay with $5.2\sigma$ significance~\cite{KsKpi}.  This decay proceeds through a $b\ra d$ penguin and a $b\ra u$ tree, so could access new physics effects.  Additionally, this channel could be used to search for an isospin partner of the $f_X(1500)$, observed in the $\Kp\Km$ spectrum of $\Bp\ra\Kp\Km\pip$~\cite{fx1500_KK}.  \babar\ measures BF$(\Bz\ra\pip\KS\Km)=(3.2\pm 0.5\pm 0.3)\ems$.  A qualitative investigation of the $\KS\Km$ spectrum reveals no evidence for an isospin partner to the $f_X(1500)$.

\section{Conclusion}

Charmless hadronic $B$ meson decays provide an excellent laboratory in which to test Standard Model predictions.  As many of these decays are governed by loop diagrams, they also present an exciting opportunity to search for new physics effects.  To date, around 100 charmless hadronic $B$ decay branching fractions have been measured with greater than $4\sigma$ significance~\cite{JimCheng}.  The physics output in charmless hadronic $B$ physics by the \babar, Belle, and CDF Collaborations continues to be strong.



\begin{thebibliography}{99}

\bibitem{JimCheng}
  H.~Y.~Cheng and J.~G.~Smith,
  Ann.\ Rev.\ Nucl.\ Part.\ Sci.\  {\bf 59}, 215 (2009)
  [arXiv:0901.4396 [hep-ph]].

\bibitem{PDG}
  K.~Nakamura {\it et al.}  [Particle Data Group],
  J.\ Phys.\ G {\bf 37}, 075021 (2010).

\bibitem{etapV} 
  P.~del Amo Sanchez {\it et al.}  [BaBar Collaboration],
  Phys.\ Rev.\  D {\bf 82}, 011502 (2010)
  [arXiv:1004.0240 [hep-ex]].

\bibitem{BelleEtapRho}
  J.~Schumann {\it et al.}  [Belle Collaboration],
  Phys.\ Rev.\  D {\bf 75}, 092002 (2007)
  [arXiv:hep-ex/0701046].

\bibitem{omegaKst}
  B.~Aubert {\it et al.}  [BABAR Collaboration],
  Phys.\ Rev.\  D {\bf 79}, 052005 (2009)
  [arXiv:0901.3703 [hep-ex]].

\bibitem{belleKstKst}
  C.~C.~Chiang {\it et al.}  [Belle collaboration],
  Phys.\ Rev.\  D {\bf 81}, 071101 (2010)
  [arXiv:1001.4595 [hep-ex]].

\bibitem{babarKstKst}
  B.~Aubert {\it et al.}  [BABAR Collaboration],
  Phys.\ Rev.\ Lett.\  {\bf 100}, 081801 (2008)
  [arXiv:0708.2248 [hep-ex]].

\bibitem{a1Kst}
  B.~Aubert {\it et al.}  [BABAR Collaboration],
  Phys.\ Rev.\  D {\bf 82}, 091101 (2010)
  [arXiv:0808.0579 [hep-ex]].

\bibitem{BelleBs}
  C.~C.~Peng {\it et al.}  [Belle Collaboration],
  Phys.\ Rev.\  D {\bf 82}, 072007 (2010)
  [arXiv:1006.5115 [hep-ex]].

\bibitem{CDFBs}
  M.~Morello  [CDF Collaboration],
  Nucl.\ Phys.\ Proc.\ Suppl.\  {\bf 170}, 39 (2007)
  [arXiv:hep-ex/0612018].
  A.~Abulencia {\it et al.}  [CDF Collaboration],
  Phys.\ Rev.\ Lett.\  {\bf 97}, 211802 (2006)
  [arXiv:hep-ex/0607021].

\bibitem{Kpi0pi0}
  P.~del Amo Sanchez {\it et al.}  [BABAR Collaboration],
  arXiv:1005.3717 [hep-ex].

\bibitem{KsKpi}
  P.~del Amo Sanchez {\it et al.}  [BABAR Collaboration],
  Phys.\ Rev.\  D {\bf 82}, 031101 (2010)
  [arXiv:1003.0640 [hep-ex]].

\bibitem{fx1500_KK}
  B.~Aubert {\it et al.}  [BABAR Collaboration],
  Phys.\ Rev.\ Lett.\  {\bf 99}, 221801 (2007)
  [arXiv:0708.0376 [hep-ex]].

\end{thebibliography}
\end{document}